\documentclass[a4paper,10pt,twoside]{cpc-hepnp}

\usepackage{multicol}
\usepackage{graphicx}
\usepackage{booktabs}
\usepackage{amssymb,bm,mathrsfs,bbm,amscd}
\usepackage[tbtags]{amsmath}
\usepackage{lastpage}
\usepackage{CJK}

\begin{document}
\begin{CJK*}{GB}{gbsn}

\fancyhead[c]{\small Chinese Physics C~~~Vol. 40, No. 8 (2016)
083102} \fancyfoot[C]{\small 083102-\thepage}

\footnotetext[0]{Received 13 January 2016, Revised 25 April 2016}

\title{The Entropy of Nonrotating Isolated Horizons in Lovelock Theory from Loop Quantum Gravity\thanks{Supported by National Natural Science
Foundation of China (11275207) }}

\author{%
     Jing-Bo Wang(王晶波)$^{1;1)}$\email{wangjb@ihep.ac.cn}%
\quad Chao-Guang Huang(黄超光)$^{1;2)}$\email{huangcg@ihep.ac.cn}
\quad Lin Li(李林)$^{2;3)}$\email{lilin@ihep.ac.cn}
}
\maketitle

\address{%
$^1$ Institute of High Energy Physics and Theoretical Physics Center for
Science Facilities, \\ Chinese Academy of Sciences, Beijing, 100049, People's Republic of China\\
$^2$ Institute of High Energy Physics, Chinese Academy of Sciences, Beijing, 100049, People's Republic of China
}

\begin{abstract}
In this paper, the BF theory method is applied to the nonrotating isolated horizons in Lovelock theory. The final entropy matches the Wald entropy formula for this theory. We also confirm the conclusion obtained by Bodendorfer et. al. that the entropy is related to the flux operator rather than the area operator in general diffeomorphic-invariant theory.
\end{abstract}

\begin{keyword}
loop quantum gravity, entropy of isolated horizons, Lovelock theory
\end{keyword}

\begin{pacs}
04.70.Dy, 04.60.Pp
\end{pacs}

\footnotetext[0]{\hspace*{-3mm}\raisebox{0.3ex}{$\scriptstyle\copyright$}2016
Chinese Physical Society and the Institute of High Energy Physics
of the Chinese Academy of Sciences and the Institute
of Modern Physics of the Chinese Academy of Sciences and IOP Publishing Ltd}%

\begin{multicols}{2}

\section{Introduction}

It has been realized for a long time that a black hole behaves like a thermal object. It has temperature and entropy \cite{bk1,hawk1}. The entropy is given by the famous Bekenstein-Hawking area law
\begin{equation}\label{0a}
    S=\frac{A}{4G\hbar},
\end{equation}
where $A$ is the area of the event horizon of a black hole.

Since the area law (\ref{0a}) is very simple, there are many methods to get this formula, see \cite{carlip1} for a review. It is difficult to tell which method or methods actually give the right explanation for the entropy of the black hole. If considering gravity theories beyond Einstein's theory, which often have black hole solutions, the entropy can be given by the Wald entropy formula \cite{wald1,wald2,wald3}. This formula has a complicated form, so cannot be obtained easily.

Lovelock theory \cite{love1} is a natural extension of general relativity in higher dimensional spacetime with higher derivative terms. This theory gives the second order Euler-Lagrange equation, so can be thought of as a toy model for ghost-free higher curvature gravity. It admits a family of AdS vacua, most (but not all) of them supporting black hole solutions \cite{love2}.

The entropy of black holes in Lovelock theory can be given by the Wald entropy formula. Can the entropy formula have a microscopic explanation? In Ref.\cite{bn1}, Bodendorfer and Neiman outline how the Wald entropy formula naturally arises in loop quantum gravity \cite{rov1,thie1,ash0,ma1} for this theory. The key observation is that in a loop quantization of a generalized gravity theory, the flux operator
turns out to measure the Wald entropy. The Chern-Simons theory they used, however, can only be defined on odd-dimensional spacetime, which is a disadvantage.

In previous papers \cite{wmz1,wh1}, the authors calculated the entropy of nonrotating isolated horizons in 4 and higher dimensional spacetime with the help of the BF theory. All those works are done in the category of pure Einstein gravity theory. In this paper, those results are extended to the Lovelock theory. In our approach it can be clearly shown that the entropy of the black hole is related to the flux operator rather than the area operator, which is just the deep insight of Ref.~\cite{bn1}.

This paper is organized as follows. In Section 2, we derive the symplectic form of the boundary theory, which can be seen as the same as the BF theory. In Section 3 a microscopic calculation of the entropy of nonrotating isolated horizons in Lovelock theory is given, and finally the Wald entropy formula is obtained. Our results are discussed in Section 4.

\section{The symplectic form}
The action of Lovelock theory is given, in $D$ dimensional spacetime, by a sum of $K=[\frac{D-1}{2}]$ terms \cite{love2},
\begin{equation}\label{1}
    I=\frac{1}{16\pi G (D-3)!}\sum_{k=0}^K \frac{c_k}{D-2 k} I_k,
 \end{equation}
which admit a compact expression in the first order formula:
\begin{equation}\label{2}
    I_k=\int \varepsilon_{a_1\cdots a_D} F^{a_1 a_2}\wedge \cdots \wedge F^{a_{2k-1} a_{2k}}\wedge e^{a_{2k+1}}\wedge\cdots \wedge e^{a_D},
\end{equation}
where $\varepsilon_{a_1\cdots a_D}$ is the anti-symmetric symbol, $F^{ab}:={\rm d}A^{ab}+A^a_{\ c}\wedge A^{cb}$ is the field strength 2-form of the spin connection 1-form $A^{ab}$, and $e^a$ is the co-vielbein 1-form. The $k=2$ term will give the usual Gauss-Bonnet term in higher dimensions.

The black hole solutions in Lovelock theory are exhaustively studied in \cite{love2}. A generalization of the event horizon of the black hole is the notion of isolated horizon \cite{ash3,ash2,abck,ash1}. The isolated horizon can also be defined in other gravity theories \cite{liko2}, and their properties  remain valid in any gravity theory, since they are of geometric origin and do not involve the field equation. In this paper, we only study the nonrotating isolated horizon.

Near the isolated horizon, we adopt the Bondi-like
coordinates $x^{\mu}=(v,r,\zeta^i)$ with coordinate indices $i,j=2,\cdots,D-1$ \cite{huang1}.  The isolated horizon $\Delta$ is characterized
by $r=0$. Following the idea of Ref. \cite{wh1} we choose a set of co-vielbein fields:
\begin{equation}\begin{split}\label{6}
    e^0_a=\sqrt{\frac{1}{2}}(\alpha  n_a +\frac{1}{\alpha}  l_a ),\,e^1_a=\sqrt{\frac{1}{2}}(\alpha  n_a -\frac{1}{\alpha} l_a ),\\
    e^\texttt{A}_a=e^\texttt{A}_{\mu}(dx^{\mu})_a ,\,\texttt{A}=2,\cdots,D-1,
\end{split}\end{equation}where $n_a,\,l_a$ are null and $e^\texttt{A}_a$ are space-like, and $\alpha(x)$ is an arbitrary function of spacetime.

On the horizon $\Delta$, the 1-form $l_a\triangleq0$, so the relevant co-vielbein satisfy
\begin{equation}\label{7}
    e^0_a\triangleq e^1_a,\quad e^\texttt{A}\triangleq \tilde{e}^\texttt{A},
\end{equation}
where $\tilde{e}^\texttt{A}$ means its value on the cross section $H$. Hereafter we denote equalities on $\Delta$ by the symbol $\triangleq$.

Defining the solder field
\begin{equation}\label{8}
    \Sigma_{IJ}=\frac{1}{(D-2)!}\varepsilon_{IJK\cdots N}e^K\wedge \cdots \wedge e^N,
\end{equation}then the non-zero solder fields on the horizon $\Delta$ satisfy
\begin{equation}\label{9}
    \Sigma_{01}\triangleq e^2\wedge e^3 \wedge \cdots \wedge e^{D-1},\,
    \Sigma_{0\texttt{A}}\triangleq -\Sigma_{1\texttt{A}}.
\end{equation}

After some straightforward calculation, the following properties of the spin connection for nonrotating isolated horizon can be obtained:
\begin{equation}\label{10}\begin{split}
    A^{01}\triangleq \kappa_l {\rm d}v+ {\rm d} (\ln\alpha(x)),\,A^{0\texttt{A}}\triangleq A^{1\texttt{A}},\,A^{\texttt{AB}}\triangleq \tilde{A}^{\texttt{AB}},
    \end{split}\end{equation}
where $\tilde{A}^{\texttt{AB}}$ is the connection comparable with the co-vielbein $\tilde{e}^\texttt{A}$.

The strength 2-form reads
\begin{equation}\label{11}
    F^{01}\triangleq 0,\quad F^{0\texttt{A}}\triangleq F^{1\texttt{A}},\quad F^{\texttt{AB}}\triangleq \tilde{F}^{\texttt{AB}},
\end{equation}
where $\tilde{F}^{\texttt{AB}}$ is the curvature of the connection $\tilde{A}^{\texttt{AB}}$.

Just like in the pure Einstein gravity theory \cite{wmz1}, the symplectic current through the isolated horizon for a single $I_k$ is
\begin{equation}\label{13}
   \int_{\Delta} J(\delta_1,\delta_2)= \int_{\Delta} \varepsilon_{a_1\cdots a_D} \delta_{[2} A^{a_1 a_2}\wedge \delta_{1]} (\tilde{F}^{a_3 a_4} \cdots \wedge \tilde{F}^{a_{2k-1} a_{2k}}\wedge e^{a_{2k+1}}\wedge\cdots \wedge e^{a_D}).
\end{equation}
Due to the properties of (\ref{7}), (\ref{10}) and (\ref{11}) on the horizon, the only term left in the above expression is
\begin{equation}\label{14}
\int_{\Delta} J(\delta_1,\delta_2)=2k(D-2)!\int_{\Delta} \delta_{[2} A^{01}\wedge \delta_{1]}(\tilde{F}^{23}\wedge \cdots \wedge \tilde{F}^{2(k-1),2k-1}\wedge e^{2k}\wedge\cdots \wedge e^{D-1}) ,
\end{equation}
where the coefficient $k$ comes from the fact that there are $k$ terms of $\tilde{F}^{ab}$, and $(D-2)!$ comes from the property of $\varepsilon_{01a_3\cdots a_D}$.

Combining all $k$ to get the full symplectic current term through $\Delta$ gives
\end{multicols}
\begin{eqnarray}\label{15}
   \int_{\Delta} J(\delta_1,\delta_2)
   =\int_{\Delta}\delta_{[2} A^{01}\wedge \delta_{1]} (\frac{1}{16\pi G (D-3)!}\sum_{k=1}^K\frac{2k(D-2)! c_k}{(D-2k)}\tilde{F}^{23}\wedge \cdots  \tilde{F}^{2(k-1),2k-1}\wedge e^{2k}\wedge\cdots \wedge e^{D-1})
=\int_{\Delta} \delta_{[2} A^{01}\wedge \delta_{1]} \pi_{01},
\end{eqnarray}
\begin{multicols}{2}
where
\begin{eqnarray}\label{16}
    \pi_{01}:=\frac{1}{16\pi G (D-3)!}\sum_{k=1}^K\frac{2k(D-2)! c_k}{(D-2k)}\tilde{F}^{23}\nonumber\\ \wedge \cdots \tilde{F}^{2(k-1),2k-1} \wedge e^{2k}\wedge\cdots \wedge e^{D-1}
\end{eqnarray}is the conjugate
momentum to the connection $A_{01}$.

For an isolated horizon,
 \begin{equation}\label{17}
    \partial_v e^\texttt{A}\triangleq0,\quad \partial_v \tilde{A}^{\texttt{AB}}\triangleq0.
\end{equation}
So $\partial_v \tilde{F}^{\texttt{AB}}\triangleq0$. Then it is easy to show that
  \begin{equation}\label{18}
    {\rm d}\pi_{01}\triangleq0,
\end{equation}
 thus it is a closed $(D-2)$-form on the horizon $\Delta$. Locally a $(D-3)$-form $B$ can be defined which satisfies
\begin{equation}\label{19}
  {\rm d}B\triangleq \pi_{01} .
\end{equation}

From Eq.~(\ref{10}) it is easy to show that
\begin{equation}\label{20}
    {\rm d}A^{01}\triangleq 0.
\end{equation}
Formulae (\ref{15}), (\ref{19}) and (\ref{20}) are similar to those in the 4 dimensional Einstein case \cite{wmz1}. Then the boundary degrees of freedom can be described by a $SO(1,1)$ BF theory isolated horizons $\Delta$, with the related fields
\begin{equation}\label{21}
    A^{01}\leftrightarrow A,\quad B\leftrightarrow B.
\end{equation}
The quantum theory for the BF theory with sources is given as in 4 dimensions: for $n$ sources $\mathcal{P}=\{p_\alpha |\alpha=1,\cdots,n\}$, the associated Hilbert space $\mathcal{H}_{BF}^n$ is spanned by basic states $|\{a_\alpha \}, a_\alpha\in \mathbb{R}>$, and the full Hilbert space is $\mathcal{H}_{BF}=\oplus_n \mathcal{H}_{BF}^n$.

\section{Calculation of the entropy}
We want to calculate the entropy of the nonrotating isolated horizon. As in pure Einstein theory \cite{wmz1}, the boundary condition can be chosen as
\begin{eqnarray}\label{23}
 {\rm d}B\circeq \pi_{01}=\frac{1}{8\pi G}(\sum_{k=1}^K\frac{k(D-2) c_k}{(D-2k)}\tilde{F}^{23}\wedge \cdots \tilde{F}^{2(k-1),2k-1}\nonumber \\ \wedge e^{2k}\wedge\cdots \wedge e^{D-1}).
\end{eqnarray}
The relative flux constraint is
\begin{eqnarray}\label{24}
    \oint_H {\rm d}B=\frac{1}{8\pi G}\oint_H \sum_{k=1}^K\frac{k(D-2) c_k}{(D-2k)}\tilde{F}^{23}\wedge \cdots \tilde{F}^{2(k-1),2k-1}\nonumber \\ \wedge e^{2k}\wedge\cdots \wedge e^{D-1}:=\tilde{a},
\end{eqnarray}

Inspired by the results obtained in loop quantum gravity in higher dimensions, the following assumption is made that the eigenvalue of the flux operator $\pi_{01}$ is quantized due to
\begin{equation}\label{25}
    \oint_H \pi_{01}(x) |{m_i,\cdots}>=\beta \sum_i m_i |{m_i,\cdots}>,\quad m_i \in \mathbb{Z}.
\end{equation}
When considering only the $k=1$ term, it returns to the result in higher dimensional pure Einstein theory.

The full Hilbert space is given by $\mathcal{H}=\mathcal{H}_V\otimes \mathcal{H}_S$, where $\mathcal{H}_V$ is the bulk Hilbert space generated by spin network states, and $\mathcal{H}_S$ is the boundary Hilbert space given by $\mathcal{H}_{BF}$ in the last section.
The quantum version of the boundary condition is
\begin{equation}\label{26}
    (\textrm{Id}\otimes \oint_S \hat{{\rm d}B}-\oint_S \hat{\pi}_{01}\otimes \textrm{Id})(\Psi_V \otimes \Psi_S)=0.
\end{equation}
After acting on the full Hilbert space, the relation between the eigenvalues of $\hat{dB}$ and $\hat{\pi}_{01}$ is
\begin{equation}\label{27}
   a_p= \beta m_p,
\end{equation}
where $a_p$ is the quantum number associated with the quantum BF theory as in the last section.

The flux constraint from the quantum version of Eq.~(\ref{24}) is
\begin{equation}\label{28}
    \sum_p |a_p|=\sum_p \beta |m_p|=\tilde{a}.
\end{equation}
or
\begin{equation}\label{29}
    \sum_p |m_p|=\frac{\tilde{a}}{\beta}:=a,\quad m_p \in \mathbb{Z}\setminus \{0\}.
\end{equation}
The eigenvalue $m=0$ drops out since it gives zero flux through the horizon, thus no contribution to the number of independent states.

Then it is easy to calculate the number of states:
\begin{equation}\label{30}
    \mathcal{N}=\sum_{n=1}^{n=a} C_{a-1}^{n-1} 2^n=2\times 3^{a-1},
\end{equation}
The entropy is given by
\begin{eqnarray}\label{31}
    S=\ln\mathcal{N}=a\ln3+\ln\frac{2}{3}=\frac{\ln3}{8\pi G\beta}\oint_H \sum_{k=1}^K\frac{k(D-2) c_k}{(D-2k)}\tilde{F}^{23}\nonumber\\ \wedge \cdots \tilde{F}^{2(k-1),2k-1} \wedge e^{2k}\wedge\cdots \wedge e^{D-1}+\ln\frac{2}{3}.
\end{eqnarray}

If $\beta=\ln3/(2\pi)$ is set, which is the same as in higher dimension pure Einstein gravity, we can get
\end{multicols}
\begin{equation}\label{31a}
    S=\frac{1}{4G}\oint_H \sum_{k=1}^K\frac{k(D-2) c_k}{(D-2k)}\tilde{F}^{23}\wedge \cdots \tilde{F}^{2(k-1),2k-1}\wedge e^{2k}\wedge\cdots \wedge e^{D-1}+\ln\frac{2}{3}=2\pi \oint_H \pi_{01}+\ln\frac{2}{3}.
\end{equation}
\begin{multicols}{2}
This is just the Wald entropy formula for Lovelock theory in terms of the flux which appears in Ref.~\cite{bn1}, plus a constant correction term. It has a simple physical picture: the entropy is just the total gravitational flux $\pi_{01}$ through the horizon section $H$ times $2\pi$.

Next let us consider a simple example: a nonrotating isolated horizon with maximal spherical symmetry, such as Schwarzschild-type black holes. In this case,
 \begin{equation}\label{32}
    F^{\texttt{AB}}\triangleq \tilde{F}^{\texttt{AB}}=\frac{1}{r_0^2} e^\texttt{A} \wedge e^\texttt{B},
\end{equation}where $r_0$ is the radius of the horizon.
Then $a$ is
\begin{eqnarray}\label{33}
    a=\frac{2\pi}{\ln 3}\oint_H {\rm d}B=\frac{1}{4G \ln 3}\oint_H \sum_{k=1}^K\frac{k(D-2) c_k}{(D-2k)r_0^{2(k-1)}}\nonumber\\ e^2\wedge \cdots \wedge e^{D-1}=\frac{A_H}{4G \ln 3}\sum_{k=1}^K\frac{k(D-2) c_k}{(D-2k)r_0^{2(k-1)}},
\end{eqnarray}
where $A_H:=\oint_H e^2\wedge \cdots \wedge e^{D-1}$ is the `area' of the horizon. The entropy is given by
\begin{equation}\label{34}
    S=a\ln3+\ln\frac{2}{3}=\frac{A_H}{4 G}\sum_{k=1}^K\frac{k(D-2) c_k}{(D-2k)r_0^{2(k-1)}}+\ln\frac{2}{3},
\end{equation}
which coincides with the result in \cite{love2,jac1} except for the constant correction term.
\section{Discussion}
In the previous sections, we studied the entropy of nonrotating isolated horizons in Lovelock theory. Finally the Wald entropy formula was obtained. Also, the parameter which appears in quantized Lovelock theory has the same value as in pure Einstein theory. This fact can be considered as a consistency check for our method, since Lovelock theory can reduce to pure Einstein theory. To get the Wald entropy formula from microscopic theory, the key assumption is the ``quantized flux" assumption (\ref{25}) in the bulk theory.

In Lovelock theory, the flux operator $\pi_{01}$, which is the conjugate momentum of the connection $A^{01}$, is not $\Sigma_{01}$ as in pure Einstein theory, but its complicated expression. This operator rather than the area operator appears in the Wald entropy formula (\ref{31a}). The $B$ field is also related to this flux operator, and through the BF theory with sources, we give the Wald entropy formula a microscopic explanation.

According to our result, the calculation of the entropy from the usual Chern-Simons theory approach in 4-dimensional spacetime should be modified. In the Chern-Simons theory approach, the area constraint plays an important role, and we think that this constraint should be replaced by the flux constraint, which is exactly the work done by Barbero et al.~\cite{blv1}. Our approach favors their results.\\

\acknowledgments{This work is supported by National Natural Science Foundation of China under the grant
11275207.}

\end{multicols}

\clearpage

\end{CJK*}

\vspace{10mm}







\begin{thebibliography}{10}

\bibitem{bk1}
J.~D. {Bekenstein},   {Phys. Rev. D,} {\bf 7}:
   2333--2346 (1973)

\bibitem{hawk1}
S.~W. {Hawking},   {Nature,} {\bf 248}:
  30--31 (1974)

\bibitem{carlip1}
S.~{Carlip},   {Lect. Notes Phys,} {\bf 769}:  89--123 (2009)

\bibitem{wald1}
R.~M. {Wald},   {Phys.
  Rev. D,} {\bf 48}:  3427--3431 (1993)

\bibitem{wald2}
V.~{Iyer} and R.~M. {Wald},   {Phys. Rev. D,} {\bf 50}:
   846--864 (1994)

\bibitem{wald3}
T.~{Jacobson}, G.~{Kang}, and R.~C. {Myers},
  {Phys. Rev. D,} {\bf 49}:  6587--6598 (1994)

\bibitem{love1}
D.~{Lovelock},   {J.
  Math. Phys.,} {\bf 12}:  498--501 (1971)

\bibitem{love2}
X.~O. {Camanho} and J.~D. {Edelstein},
  {Class. Quant. Grav.,} {\bf 30}:  035009 (2013)

\bibitem{bn1}
N.~{Bodendorfer} and Y.~{Neiman},  {Phys. Rev. D,} {\bf 90}:  084054 (2014)

\bibitem{rov1}
C.~Rovelli, {\em Quantum Gravity}.
(\newblock Cambridge Monographs on Mathematical Physics. Cambridge University
  Press, 2004)

\bibitem{thie1}
T.~Thiemann, {\em Modern Canonical Quantum General Relativity}.
(\newblock Cambridge Monographs on Mathematical Physics. Cambridge University
  Press, 2008)

\bibitem{ash0}
A.~{Ashtekar} and J.~{Lewandowski},  {Class. Quant. Grav.,} {\bf 21}:  R53 (2004)

\bibitem{ma1}
M.~{Han}, W.~{Huang}, and Y.~{Ma},   {Int. J. Mod. Phys. D,} {\bf 16}:  1397--1474 (2007)


\bibitem{wmz1}
J.~{Wang}, Y.~{Ma}, and X.-A. {Zhao},   {Phys. Rev. D,} {\bf 89}:
   084065 (2014)

\bibitem{wh1}
J.~{Wang} and C.-G. {Huang},  {Class. Quant. Grav.,}
  {\bf 32}:  035026 (2015)

\bibitem{ash3}
A.~{Ashtekar}, C.~{Beetle}, and S.~{Fairhurst},  {Class. Quant. Grav.,} {\bf 16}:
   L1--L7 (1999)

\bibitem{ash2}
A.~{Ashtekar}, S.~{Fairhurst}, and B.~{Krishnan},  {Phys. Rev. D,} {\bf 62}:
  104025 (2000)

\bibitem{abck}
A.~{Ashtekar}, J.~C. {Baez}, A.~{Corichi}, and K.~Krasnov,  {Phys. Rev. Lett.,} {\bf 80}:
  904--907 (1998)

\bibitem{ash1}
A.~{Ashtekar}, J.~C. {Baez}, and K.~{Krasnov}, {Adv. Theor. Math. Phys.,}
  {\bf 4}:  1--94 (2000)

\bibitem{liko2}
T.~{Liko} and I.~{Booth},  {Class. Quant.Grav.,} {\bf 24}:
  3769--3782 (2007)

\bibitem{huang1}
X.-N. {Wu}, C.-G. {Huang}, and J.-R. {Sun},   {Phys. Rev. D,} {\bf 77}:
   124023 (2008)

\bibitem{jac1}
T.~{Jacobson} and R.~C. {Myers},  {Phys. Rev. Lett.,} {\bf 70}:  3684--3687 (1993)

\bibitem{blv1}
G.~F.~{Barbero}, J.~{Lewandowski}, and E.~{Villasenor},  {Phys. Rev. D,} {\bf 80}:  044016 (2009)

\end{thebibliography}
\end{document}